\documentclass[11pt,twocolumn,twoside]{osajnl}
\usepackage{mathtools}
\usepackage{amsmath}
\usepackage{units}
\usepackage{upgreek}
\graphicspath{ {./image/} }
\DeclarePairedDelimiter\abs{\lvert}{\rvert}

\journal{ao} 

\setboolean{shortarticle}{true}

\newcommand{\rvec}{\mathbf{r}} 
\newcommand{\e}{{\rm e}} 
\renewcommand{\i}{{\rm i}} 
\newcommand{\nmax}{n_\text{max}} 
\newcommand{\ym}{\mathbf{y}_\text{m}}
\newcommand{\yr}{\mathbf{y}_\text{r}}
\renewcommand{\wr}{\mathbf{w}_\text{r}}
\newcommand{\X}{\mathbf{X}} 
\newcommand{\A}{\mathbf{A}} 
\newcommand{\B}{\mathbf{B}} 
\newcommand{\C}{\mathbf{C}} 
\newcommand{\NA}{\textrm{NA}} 
\newcommand{\clim}{c_\text{lim}} 

\title{Focal field analysis of highly multi-mode fiber beams based on modal decomposition}

\author[1,2,*]{Hao Pang}
\author[1]{Tobias Haecker}
\author[3]{Alexandre Bense}
\author[2]{Tobias Haist}
\author[4]{Daniel Flamm}

\affil[1]{TRUMPF Werkzeugmaschinen GmbH + Co. KG, Johann-Maus-Strasse 2, 71254 Ditzingen, Germany}
\affil[2]{Institut of Applied Optics, University of Stuttgart, Pfaffenwaldring 9, 70569 Stuttgart, Germany}
\affil[3]{ECOLE POLYTECHNIQUE, PALAISEAU Cedex, 91128, France}
\affil[4]{TRUMPF Laser-und Systemtechnik GmbH, Johann-Maus-Strasse 2, 71254 Ditzingen, Germany}

\affil[*]{hao.pang@trumpf.com}

\ociscodes{(140.3295) Laser beam characterization, (140.3300) Laser beam shaping, (140.3390) Laser materials processing, (060.2270) Fiber characterization, (030.4070) Modes. }

\doi{\url{https://doi.org/10.1364/AO.397498}}

\begin{abstract}
 In this work, a numerical modal decomposition approach is applied to model the optical field of laser light after propagating through a highly multi-mode fiber. The algorithm for the decomposition is based on the reconstruction of measured intensity profiles along the laser beam caustic with consideration of intermodal degrees of coherence derived from spectral analysis. To enhance the accuracy of the model, different approaches and strategies are applied and discussed. The presented decomposition into a set of LP modes enables both the wave-optical simulation of radiation transport by highly multi-mode fibers and, additionally, the analysis of free-space propagation with arbitrarily modified complex amplitude distributions.
\end{abstract}

\setboolean{displaycopyright}{false}

\begin{document}

\maketitle
\\
\textbf{
\copyright\,\,2020 Optical Society of America. One print or electronic copy may be made for personal use only. Systematic reproduction and distribution, duplication of any material in this paper for a fee or for commercial purposes, or modifications of the content of this paper are prohibited.}

\section{Introduction}\label{section:1}
The high power solid-state laser is a widely spread instrument for industrial material processing. The global market for laser material processing systems reached $\unit[\$19.8]{Billion}$ in 2018 \cite{lasermarket1}. Considering laser processing strategies of recent years, a shift from directly increasing laser powers to smartly and efficiently exploit the power performance of the laser platforms has been observed. Here, remarkable examples can be found in the multi-spot technology for high-quality welding of zinc-coated steel \cite{reimann2017three} or the use of Bessel-like beams for efficient cutting of transparent materials \cite{jenne2018high}. Such beam shaping concepts show promising progress in improving the quality and increasing the throughput of established laser application processes, or enable the development of completely new application strategies \cite{OlsenFl, goppold2014experimental, flamm2019beam}.\\
Because of its flexibility and rather low expense, the standard industrial beam transportation solution from a solid-state laser (in this paper operating at $\lambda\approx\unit[1]{\upmu m}$) into a processing unit is a step-index fiber \cite{hugel2009laser}. To deliver high-optical average powers in the multi-kW regime for applications such as laser cutting or welding, the dimensions of the light guiding core need to be extended to minimize non-linear effects. These multi-mode fibers can transmit radiation in a wide range of spatial quality from $M^2 \approx 1$, thus, close to the diffraction limit to $M^2 >> 10$ with thousands of propagating modes and complex coherence situations. \\
To get an accurate wave-optical description for multi-mode beams, modal decomposition techniques have been implemented to model the laser source \cite{Wolf:82, Wolf:86}. By solving the Helmholtz equation with a given refractive index distribution of a weakly guiding step-index fiber (i.e. with a small \NA), a set of linearly polarized (LP) modes is obtained \cite{snyder2012optical}. Additionally, the modal degree of coherence determined by the spectrum of the light source and its impact to the intensity profile is introduced and discussed. For this purpose we consider in detail optical path length differences of individual modes inside the fiber and their ability to interfere \cite{flamm2013modal}.\\
The aim of a modal decomposition is to access the modal power spectra to completely describe complex field distributions with few mode coefficients only \cite{shapira2005complete, kaiser2009complete}. Shapira \textit{et al.}\@ \cite{shapira2005complete} and Bruening \textit{et al.}\@ \cite{bruning2013comparative} have demonstrated a numerical approach for modal decomposition based on intensity profiles in near- and far-field. Alternatively, direct experimental access to modal power spectra is provided by using, e.g., correlation filters \cite{kaiser2009complete, forbes2016creation} (and references therein) or $S^2$ imaging techniques \cite{nicholson2008spatially}. Remarkably, with the latter approach, besides the modal power spectrum, the spatial field distribution of propagating modes can be reconstructed \cite{nicholson2008spatially}. However, all mentioned methods have only been applied to few-mode fibers and, considering the number of investigated modes, cannot be scaled easily. In the present application, on the other hand, multi-mode waveguides with relatively large $V$-parameters are investigated, which allow the stable propagation of over several hundreds of LP-modes. Our main focus is not to observe the exact modal weight information, but to find the general tendency of the excited modal weight spectrum, to accurately model the emerging radiation in a wave-optical manner. This should include interference effects caused by the (partial) degree of coherence of the source. For this purpose, a numerical approach is presented here that evaluates intensity profiles taken from standardized caustic measurements.\\
The determination of the modal power spectrum of a highly multi-mode fiber contributes not only to the exact focal field analysis required for beam shaping \cite{tillkorn2018anamorphic}, but also to research fields such as fiber characterization, imaging technologies \cite{flusberg2005fiber,borhani2018learning,ploschner2015seeing} and optical communication  \cite{jacobs2002optical}.\\
The paper is organized as follows. In Sec.\,\ref{section:2} the intermodal degree of coherence based on a spectral analysis is introduced. Afterwards, in Sec.\,\ref{section:3} we show that the resulting problem can be solved by a linear least square regression. Section.\,\ref{section:4} treats potential error sources from experimentally obtained data and their impact to the decomposition. Finally, in Sec.\,\ref{section:5} we try to give an explanation why numerical mode analysis techniques based on intensity measurements suffer from ambiguities or inaccurate reconstruction results and introduce the concept of modal diversity.\\
\section{mode superposition and the modal degree of coherence} \label{section:2}
\begin{figure}[t]
    \centering
    \includegraphics[width=\linewidth]{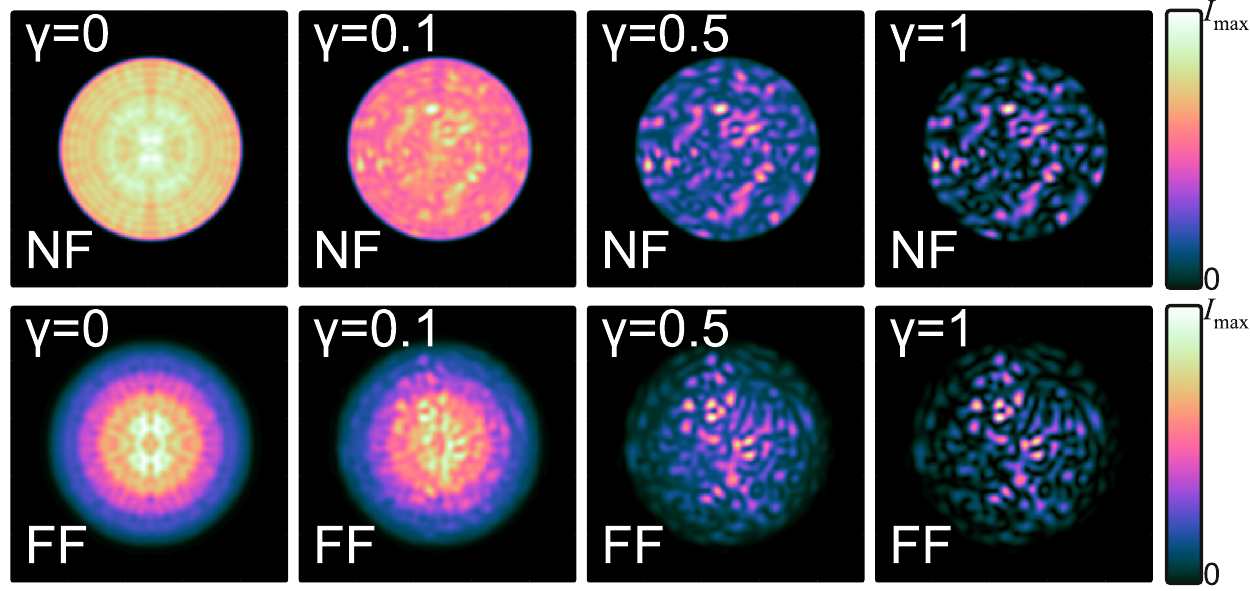}
    \caption{Synthetic transverse intensity profiles considering different modal degrees of coherence with overall $\abs{\gamma}=0$ (incoherent), $0.1, 0.5$ (partial coherent), $1$ (coherent). Near-field (top) and far-field (bottom) intensity profiles are depicted from a composition of 328 modes. While $\gamma \approx 0$ is close to the situation for the high power experiments, cf.\,Sec.\,\ref{section:4}\,\ref{sec:selectedExample}, speckle-like distributions for $\abs{\gamma} \approx 1$ are known from, e.g., endoscopic concepts using multi-mode fibers \cite{ploschner2015seeing}.}
    \label{fig:coherence_nf_ff}
\end{figure}

We consider a scalar optical field $U(\rvec)$ represented as superposition of eigenmodes $\Psi_n(\rvec)$ with modal weights $\varrho_n \left( n = 1 \ldots \nmax \right)$. The resulting intensity profile reads
\begin{align}\label{eq:1}
    I\textsubscript{total}(\rvec)&=U(\rvec)U^*(\rvec) \nonumber\\
                &=\sum_{m,n}\varrho_m\varrho_n\abs{\gamma_{mn}}\abs{\Psi_m}\abs{\Psi_n} \e^{\i\left(\Delta\phi_{mn} + \Delta\chi_{mn}(\rvec)\right)},
\end{align} 
where $\Delta\phi_{mn}$ is the intermodal phase difference and the spatial phase offset $\Delta\chi_{mn}$ results from the different local phase relations, which is related to the spatial coordinate $\rvec$. The complex valued degree of temporal coherence between two modes is defined as $\gamma_{mn}$ \cite{otto2013improved}. As known from statistical optics, we can split this sum into an incoherent term $I\textsubscript{inc}\left(\mathbf{r}\right)$ and an interference term $I\textsubscript{int}\left(\mathbf{r}\right)$, i.e.
\begin{align}\label{eq:2}
    &I\textsubscript{total}(\rvec)=I\textsubscript{inc}(\rvec)+I\textsubscript{int}(\rvec) \nonumber\\
    &=\sum_{n}^{\nmax}{\varrho_n^2\abs{\Psi_n}^2} \nonumber\\
    &+2\sum_{\substack{m,n\\m>n}}^{\nmax} {\varrho_m\varrho_n\abs{\gamma_{mn}}\abs{\Psi_m}\abs{\Psi_n}\cos\left[\Delta\phi_{mn}+\Delta\chi_{mn}(\rvec)\right]}.
\end{align}
Thus, if a complete modal decomposition \cite{kaiser2009complete, shapira2005complete} is performed with $\nmax$ modes, $\left(\nmax^2-\nmax\right)/2$ interference terms have to be considered.\\
The temporal coherence $\gamma_{mn}$, which weights the phase dependent cosine modulation in the interference term $I\textsubscript{int}(\rvec)$, can be estimated appropriately by calculating the mode dependent optical path length. According to Wiener-Khinchin theorem and assuming that each mode shares the same spectral information, $\gamma$ is proportional to the Fourier transform of light's spectral density $S\left(\nu\right)$ \cite{cohen1998generalization,born2013principles,flamm2013modal}:
\begin{equation}\label{eq:3}
    \gamma(\Delta L)=\abs*{\frac{\int_{0}^{\infty} \text{d}\nu S(\nu)\e^{-2\pi \i \nu \Delta L/c}}{\int_{0}^{\infty} \text{d} \nu S(\nu)}}.
\end{equation}
Here, $\nu$ is the frequency of light and $\Delta L$ is the optical path difference which depends on the difference between effective refractive index $\Delta n\textsubscript{eff}$ for each mode pair. Denoting the propagation constant of the respective mode by $\beta_m$, the effective refractive index $n\textsubscript{eff}$ can be expressed by 
\begin{equation}\label{eq:4}
n_{\text{eff},m}=\beta_m\frac{\lambda}{2\pi}. 
\end{equation}
After travelling through a fiber of length $L$, the modal optical path difference is given by $\Delta L_{mn}= \Delta n_{\text{eff},mn}L$.\\
\begin{figure}[h]
\centering
\includegraphics[width=\linewidth]{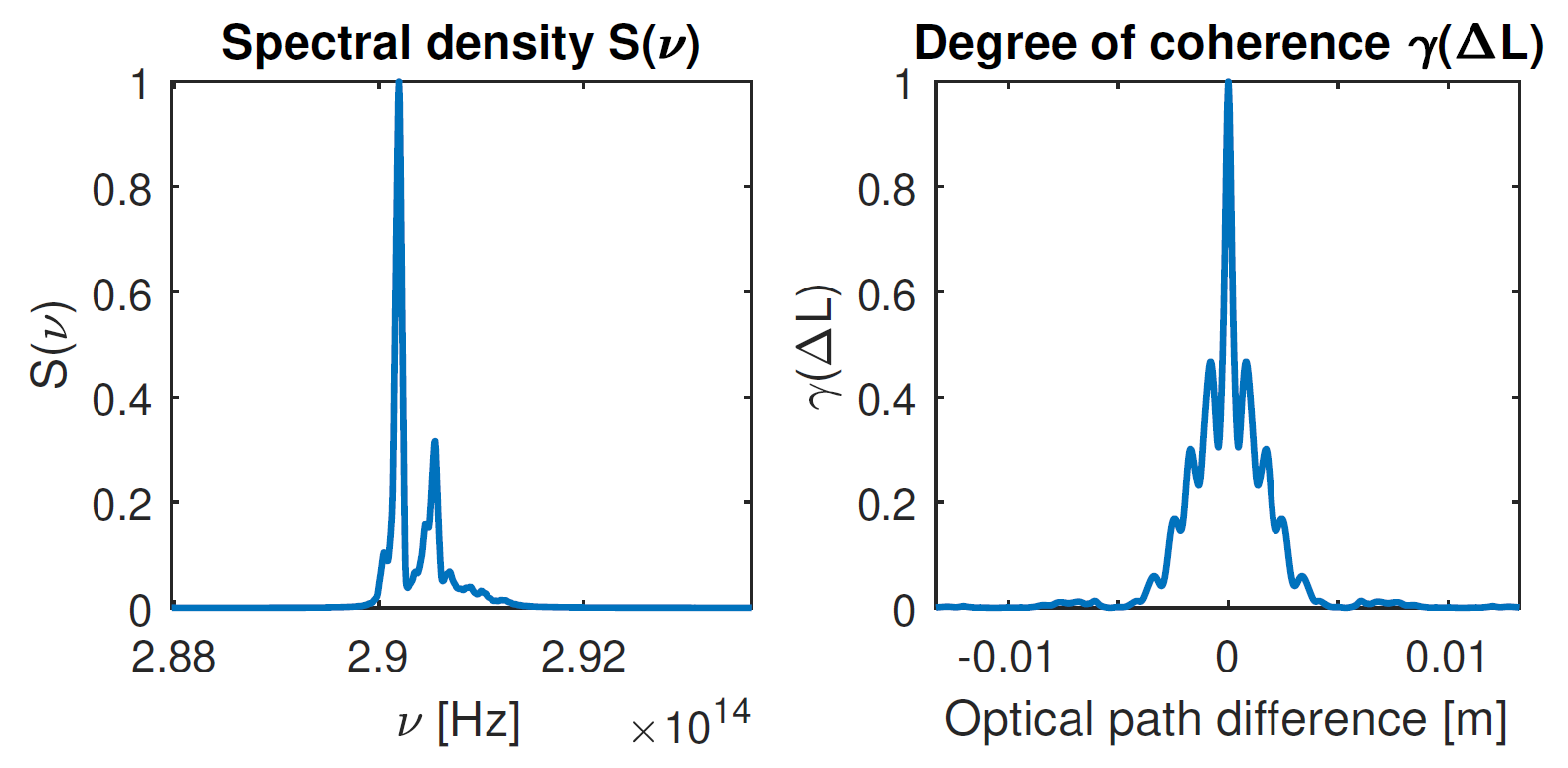}
\caption{Left: spectral density of a single mode diode laser ($\lambda=\unit[1030]{nm}$). Right: calculated degree of coherence depending on modal optical path difference.}
\label{fig:spectrum}
\end{figure}
Consequently, a separate modal degree of coherence for each mode pair can be found as $\gamma_{mn}$, which varies between 0 and 1. If $\abs{\gamma_{mn}}=1$, the superposition of two modal fields is considered fully coherent. On the contrary, if $\abs{\gamma_{mn}}=0$, the interference term is zero, which corresponds to a complete incoherent superposition. If $0<\abs{\gamma_{mn}}<1$, the partial coherence is considered \cite{flamm2013modal}. In Fig.\,\ref{fig:spectrum}, the spectrum of a diode laser source with $\lambda$ = \unit[1030]{nm} and the corresponding temporal degree of coherence according to \eqref{eq:3} is presented. For a  rough estimation of the coherence length $L\textsubscript{c}$, the well-known approximation of Gaussian spectra is applied $L\textsubscript{c}\approx {\lambda^2} / {\Delta \lambda}={(\unit[1030]{nm})^2} / {\unit[0.93]{nm}}\approx\unit[1.1]{mm}$ \cite{born2013principles}, cf.\,Fig.\,\ref{fig:spectrum}. Thus, in a very good approximation, if $\Delta L_{mn} > L_\textsubscript{c}$, the modal degree of coherence $\gamma_{mn}\approx 0$ and no interference is observed. The depicted spectrum is representative for the laser sources employed for the high power measurements discussed in Sec.\,\ref{section:4}. Additionally, examples of simulated intensity profiles with consideration of different modal degrees of coherence are depicted in Fig.\,\ref{fig:coherence_nf_ff}. Here, increasing interference contrast of the speckle-like intensity profiles with increasing modal degree of coherence becomes clearly visible. In all four cases, for the sake of convenience, all corresponding mode pairs exhibit the same degree of coherence, which can be estimated by the well-known fringe visibility $\abs{\gamma} \approx V = \left(I_{\text{max}} - I_{\text{min}}\right) / \left(I_{\text{max}} + I_{\text{min}}\right)$ \cite{born2013principles}. By means of the selected examples discussed in Secs.\,\ref{section:3} and \ref{section:4}\,\ref{sec:selectedExample}, respectively, we will show the impact of a partial coherent mode composition for an accurate focal field description.\\
In general, the laser radiation to be coupled into the fiber exhibits diverse longitudinal and transverse modes \cite{hodgson2005laser}. Those modes excite their own modal weight distributions with slightly different frequency spectra during the coupling and the propagation within the fiber, which makes the coherence situation even more complex and in general reduces the modal degree of coherence ($\gamma_{mn}\textsubscript{, real} < \gamma_{mn}$). For convenience, all modes are considered to share the same spectrum.
\section{linear least square solution}\label{section:3}
\begin{figure*}[h]
    \centering
    \includegraphics[width=\textwidth]{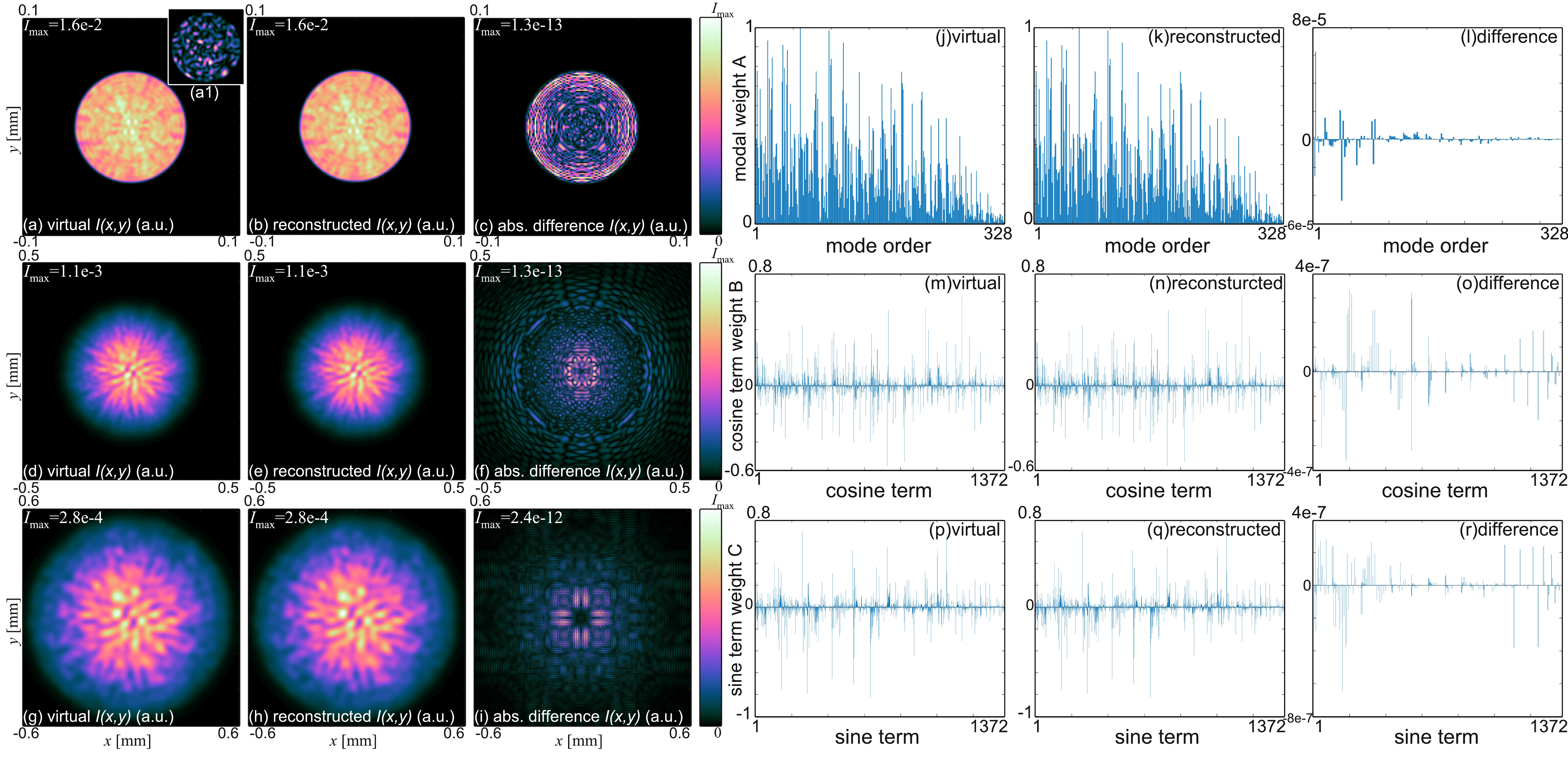}
    \caption{Modal decomposition based on synthetic intensity distributions. Simulated synthetic transverse intensity profiles at different propagation distances $z=0$ (correspond to fiber end facet) (a), $z=\unit[2.03]{mm}$ (d), and $z=\unit[4.7]{mm}$ (g) with corresponding reconstructions (b), (e), and (g). Absolute intensity differences between synthetic and reconstructed data for each propagation distance is depicted in (c), (f), and (i). The synthetic modal weights, its simulated reconstruction and the corresponding differences are depicted in (j), (k), and (l), respectively. Further plots (m), (n), (o), (p), (q), and (r) present the modal weight information of cosine and sine terms, respectively, cf.\,Eq.\,(\ref{eq:8}). 2744 (1372 cosine and 1372 sine) interference terms were considered in both synthetic intensity distributions and reconstructions. The number of totally considered terms is 3072. The inset (a1) shows an example with full coherence $\abs{\gamma}=1$ and high interference contrast.
    }
    \label{fig:intensityVI24_weightVI24}
\end{figure*}

For a complete modal decomposition, the optimization problem can be reformulated as a linear least square problem. Firstly, we split the interference term in \eqref{eq:2} into a sine and a cosine term, i.e.: 
\begin{align}
    I_{\text{int}}^{\text{cos}}=&2\sum_{\substack{m,n\\m> n}}^{\nmax} \varrho_m\varrho_n\abs{\gamma_{mn}}\abs{\Psi_m}\abs{\Psi_n}\cos\left(\Delta\phi_{mn}\right) \nonumber\\
    &\times\cos\left[\Delta\chi_{mn}(\rvec)\right],    \label{eq:6}   \\
    I_{\text{int}}^{\text{sin}}=&-2\sum_{\substack{m,n\\m> n}}^{\nmax} \varrho_m\varrho_n\abs{\gamma_{mn}}\abs{\Psi_m}\abs{\Psi_n}\sin\left(\Delta\phi_{mn}\right) \nonumber\\ 
    &\times\sin\left[\Delta\chi_{mn}(\rvec)\right].    \label{eq:7}
\end{align}
with the introduced new weight factors
\begin{align}\label{eq:8}
    \mathbf{A}&=\varrho_n^2, \nonumber \\
    \mathbf{B}&=\varrho_m\varrho_n\abs{\gamma_{mn}}\cos\left(\Delta\phi_{mn}\right), \nonumber \\
    \mathbf{C}&=\varrho_m\varrho_n\abs{\gamma_{mn}}\sin\left(\Delta\phi_{mn}\right),
\end{align}
the complete intensity $I\textsubscript{total}(\rvec)$ is rewritten as
\begin{align}\label{eq:9}
    I\textsubscript{total}(\rvec)=&\sum_{n}^{\nmax}
    {\A\abs{\Psi_n}^2}\nonumber \\
    +&2\sum_{\substack{m,n\\m> n}}^{\nmax} \B\abs{\Psi_m}\abs{\Psi_n}\cos\left[\Delta\chi_{mn}(\rvec)\right] \nonumber\\
    -&2\sum_{\substack{m,n\\m> n}}^{\nmax} \C\abs{\Psi_m}\abs{\Psi_n}\sin\left[\Delta\chi_{mn}(\rvec)\right]
\end{align}
with the summation of $\A$ is one. $\A, \B$ and $\C$ are vectors. The ansatz functions are defined as the corresponding terms $\abs{\Psi_n}^2$, $2\abs{\Psi_m}\abs{\Psi_n}\cos\left[\Delta\chi_{mn}(\rvec)\right]$ and $2\abs{\Psi_m}\abs{\Psi_n}\sin\left[\Delta\chi_{mn}(\rvec)\right]$ of each weight factor.
Thus, in total $t = \nmax + 2\left[(\nmax^2-\nmax)/2\right] = \nmax^2$ parameters merged in
\begin{equation}
    \mathbf{w} = (\A, \B, \C)^\top
\end{equation}
are unknown. we assume that the intensity profile is measured on rectangular $(x, y)$-grids in $N$ distinct planes for the reconstruction. The measurements are stored in a column vector with structure
\begin{equation}\label{eq:9.1}
    \mathbf{y} = \left(I\left(x,y; 1\right), \ldots, I\left(x,y;N\right)\right)^\top.
\end{equation}
Then, for discrete intensity data, \eqref{eq:9} can be rewritten as a matrix-vector product
\begin{equation}\label{eq:10}
    \mathbf{y}=\mathbf{X}\mathbf{w},
\end{equation}
with $\A$,$\B$ and $\C$ are the weight factors in $\mathbf{w}$. The measured intensity distributions are stored in $\mathbf{y}$, see \eqref{eq:9.1} and the corresponding ansatz function terms are reformulated in $\X$. The structure of $\X$ is
\begin{equation}\label{eq:11}
\mathbf{X}=
    \begin{pmatrix}
    I_{1,1}^{1,1}&\cdots&I_{1,1}^{t,1}\\
    \vdots&\ddots&\vdots\\
    I_{x,y}^{1,N}&\cdots&I_{x,y}^{t,N}\\
    \end{pmatrix}.
\end{equation}
Thus, each column represents the evaluation of one of the $t$ ansatz functions sorted by the $N$ measured planes along the caustic.\\
Denoting now the measured data by $\ym$ and the reconstructed intensity by
\begin{equation}
    \yr = \X \wr,
\end{equation}
the squared residuum reads 
\begin{equation}\label{eq:12}
    \Delta^N = \|\ym - \X\wr \|^2.
\end{equation}
This residuum is minimized in a linear least square regression by solving 
\begin{equation}\label{eq:13}
    \X^\top \X\wr = \X^\top\ym.
\end{equation}
In case that $\X$ has full column rank, i.e. the columns of $\X$ are linearly independent, the solution for $\wr$ is unique and can be calculated by
\begin{equation}\label{eq:14}
    \wr=(\X^\top \X)^{-1}\X^\top\ym.
\end{equation}
This condition is checked numerically by calculating $\det(\X^\top\X)$ for a given set of ansatz functions.\\
Usually, the $\NA$ of the investigated laser beam is considerably smaller than the $\NA$ of the optical fiber into which the radiation of the original laser is coupled into. Thus, a pre-selection of modes is taken into account for the reconstruction based on their individual (modal) $\NA$ to reduce the computing effort. In our case, 328 of 1148 modes remained after the pre-selection ($\NA = 0.22$ and $r\textsubscript{core}=\unit[50]{\upmu m}$, $ L\textsubscript{fiber}=\unit[30]{m}$, cf.\,Sec.\,\ref{section:4}\,\ref{sec:section4D}).\\
To test the decomposition method, it is firstly applied to a set of synthetic intensity distributions with predetermined modal weights and phase differences. Here, partial coherence is considered by calculating the degree of coherence according to Sec.\,\ref{section:2} and the applied coherence information is comparable to $\gamma(\Delta L)$ from Fig.\,\ref{fig:spectrum}. The synthetic intensity profiles are depicted on the left-hand side with corresponding modal weight distributions shown on the right-hand side of Fig.\,\ref{fig:intensityVI24_weightVI24}, respectively. To reduce the computational effort, this synthetic test beam is composed of 1372 interference terms. Additionally, an intensity profile with full coherence is presented in (a1) of Fig.\,\ref{fig:intensityVI24_weightVI24}. Comparing (a1) and (a) of Fig.\,\ref{fig:intensityVI24_weightVI24}, a significant increase of contrast can be observed, which indicates the impact of $\gamma_{mn}$ and the necessity to consider it for a precise focal field description, cf.\,Sec.\,\ref{section:2}. The entire calculation was based on a conventional computer (128 GB RAM) and took only a few minutes.\\
The absolute difference between synthetic data and the reconstructed intensity distributions depicted in (c), (f) and (i) of Fig.\,\ref{fig:intensityVI24_weightVI24} is 10 orders of magnitude smaller than the input intensity. Moreover, the difference between the input weight factors and the reconstructed ones shown in (l), (o) and (r) of Fig.\,\ref{fig:intensityVI24_weightVI24} is also about 5 orders of magnitude smaller than the specified modal weights. Thus, in the situation where the data for the reconstruction is in the image space of the ansatz functions, the presented linear least square regression works excellently.
\\
\section{Experimental Implementation}\label{section:4}
Similar to the synthetic data, cf.\,Fig.\,\ref{fig:intensityVI24_weightVI24}, we recorded experimental data $\ym$ via caustic measurements according to ISO 11146 \cite{ISO11146}. A standardized caustic measurement usually takes a few minutes. During this measurement time, the mode set under investigation should remain constant in first approximation. 
In case of real measurement data for the spatial intensity distribution $\ym$, camera noise and aberrations are unavoidable. Those disturbances make the implementation of modal decomposition described in Sec.\,\ref{section:3} inaccurate and deficient. The ansatz function matrix $\X$, the measured data $\ym$ and the numerical method must be adapted to the real application environment, so that the least square residuum introduced in Sec.\,\ref{section:3} is still applicable. This adaption is discussed in the following.
\subsection{Camera noise and dynamic range}\label{sec:section:4A}
Camera noise can directly be considered as deviation in the measurement data $\ym$. To mitigate the deviation, the mean value of repeated measurements is taken to improve the signal-to-noise ratio (SNR). Additionally, background map subtraction is applied, as known from standards for camera-based laser beam characterization methods, see ISO 11146 \cite{ISO11146}. \\ Although not performed in this study, we refer to \cite{merx2020TBS} for the beneficial use of high dynamic range (HDR) intensity recordings of beam profiles for improving the dynamical range and for noise reduction.
\subsection{Aberration of imaging system}
\begin{figure}[t]
    \centering
    \includegraphics[width=\linewidth]{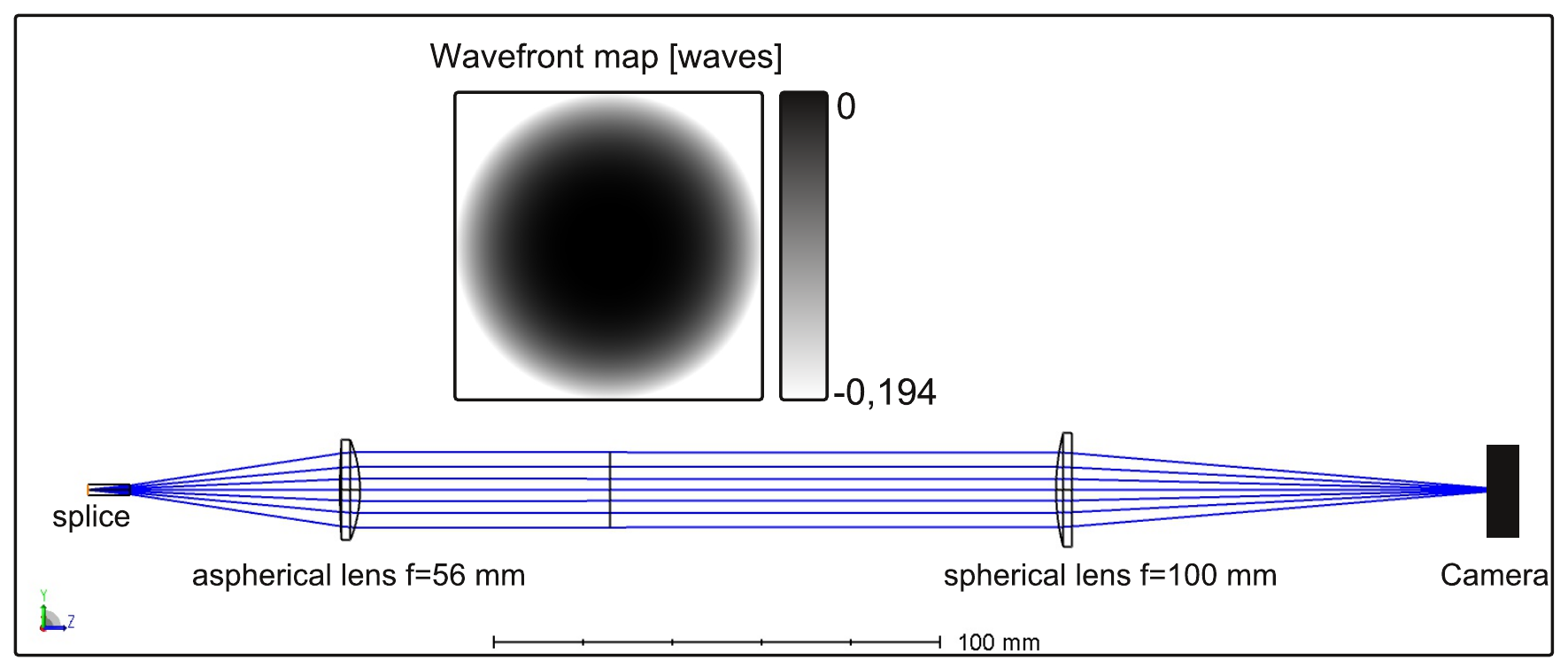}
    \caption{Calculation of aberration based on ray tracing for a typical imaging system. Left lens: aspherical, $f=\unit[56]{mm}$; right lens: spherical, $f=\unit[100]{mm}$. Extra mirrors for attenuation of laser power are not presented. A polarizer is optional if a polarisation-resolved measurement is needed.}
    \label{fig:aberration}
\end{figure}

Since a real optical imaging system is never perfect, aberrations remain existing in each intensity measurement. We analyse aberrations occurring during propagation through our processing optical setup using geometrical optics, see example depicted in Fig.\,\ref{fig:aberration}, and then considered in the ansatz function matrix $\X$.
Generally, Zernike polynomials are applied to describe wave front aberrations of an imaging system. The aberrated mode field $\Psi\textsubscript{abrr}\left(\rvec\right)$ can be calculated by multiplying the original mode field $\Psi\left(\rvec\right)$ with the wavefront aberration $W\left(\rvec'\right)$ in the far field and transforming it back to the near field, i.e. $\Psi\textsubscript{abrr}(\rvec) = \mathcal{F}^{-1}\left\{\mathcal{F}\left[\Psi(\rvec)\right]W(\rvec')\right\}.$
Due to the small extent of the field, the wavefront aberration is typically dominated by spherical aberration. Since different modes exhibit different divergent angles, the beam waist shift of the reconstructed laser beam depends directly on the reconstructed modal weight. Thus, the Fourier transformed field $\Psi\textsubscript{abrr}\left(\rvec\right)$ might not be exactly at the beam waist position. A correction of beam waist position is therefore necessary, which also led us to further develop the method and to implement an iterative procedure, cf.\, Sec.\,\ref{section:4}\,\ref{ss:im}. \\
The imaging quality of the processing optics analyzed here using ray tracing, cf.\,Fig.\,\ref{fig:aberration}, may also be determined experimentally in various ways. As an example, we refer to Merx \textit{et al.}\ \cite{merx2020TBS} where phase profiles of coherent radiation are reconstructed from solving the transport-of-intensity equation. Equally to the concept for modal decomposition presented here, Ref.\,\cite{merx2020TBS} makes use of caustic measurements forming the basis for a laser beam characterization.
\subsection{Spatial mismatch of coordinate systems}\label{sec:section4C}
The center of the calculated LP modes always corresponds to the center of the optical fiber and the optical axis. For real intensity measurements, the optical axis is usually not defined and is determined by the center of the measured intensity profile, which is calculated by its first-order moment \cite{ISO11146}. However, due to camera noise, aberrations and interference, the first-order moment of the intensity distribution does not necessarily match the optical axis of the laser beam. With the caustic measurements performed according to ISO 11146, there are at least 20 layers in near and far field available \cite{ISO11146}. A linear fit of the propagation dependent first-order moments based on multiple layers yields accurate information about the optical axis.
\subsection{Neglect of interference terms}\label{sec:section4D}
In case of a fiber with $\NA = 0.22$ and $r\textsubscript{core}=\unit[50]{\upmu m}$, there are $\nmax=1148$ modes and thus about $650000$ interference terms. This fiber is a representative waveguide for guiding multi-kW optical powers to a processing unit for e.g. cutting of sheet steel. After a pre-selection mentioned in Section \ref{section:3}, there are $\nmax=328$ modes and $53628$ remaining interference terms. Due to limited calculation power, it is necessary to reduce the dimension of matrix $\X$, which means that not all interference terms are considered. A selection of significant interference terms is based on modal weights and the estimated intermodal degree of coherence, cf.\,Sec.\,\ref{section:2}. For this purpose, an appropriate threshold value $\clim$ is chosen and only interference terms for which the condition 
\begin{equation}\label{eq:filter}
    \abs{\gamma_{mn}}\varrho_m\varrho_n > \clim
\end{equation}
is met are considered in the subsequent partially coherent superposition. Interference terms of certain mode pairs with low modal weights $\rho_n\rho_m$ and low modal degrees of coherence $\gamma_{mn}$ are therefore neglected. The intermodal degree of coherence is estimated by spectral information as stated in \eqref{eq:3} and the $\clim$ depends directly on the calculation power and the computer memory. We take an assumption that all interference terms satisfying $\rho_m\leqslant1\%$, $\rho_n\leqslant1\%$ and $\abs{\gamma_{mn}}\leqslant0.01$ are neglected, thus $\clim = 10^{-6}$. Another option to define $\clim$ is to take the coherence length $L_\textsubscript{c}$ into consideration, cf.\,Sec.\,\ref{section:2}.\\
The selection of interference terms reduces massively the dimension (number of columns) of $\X$ and is also an essential step in the iterative method described in the following Sec.\,\ref{section:4}\,\ref{ss:im}.
\subsection{Iterative Method}\label{ss:im}
\begin{figure}[t]
    \centering
    \includegraphics[width=0.775\linewidth]{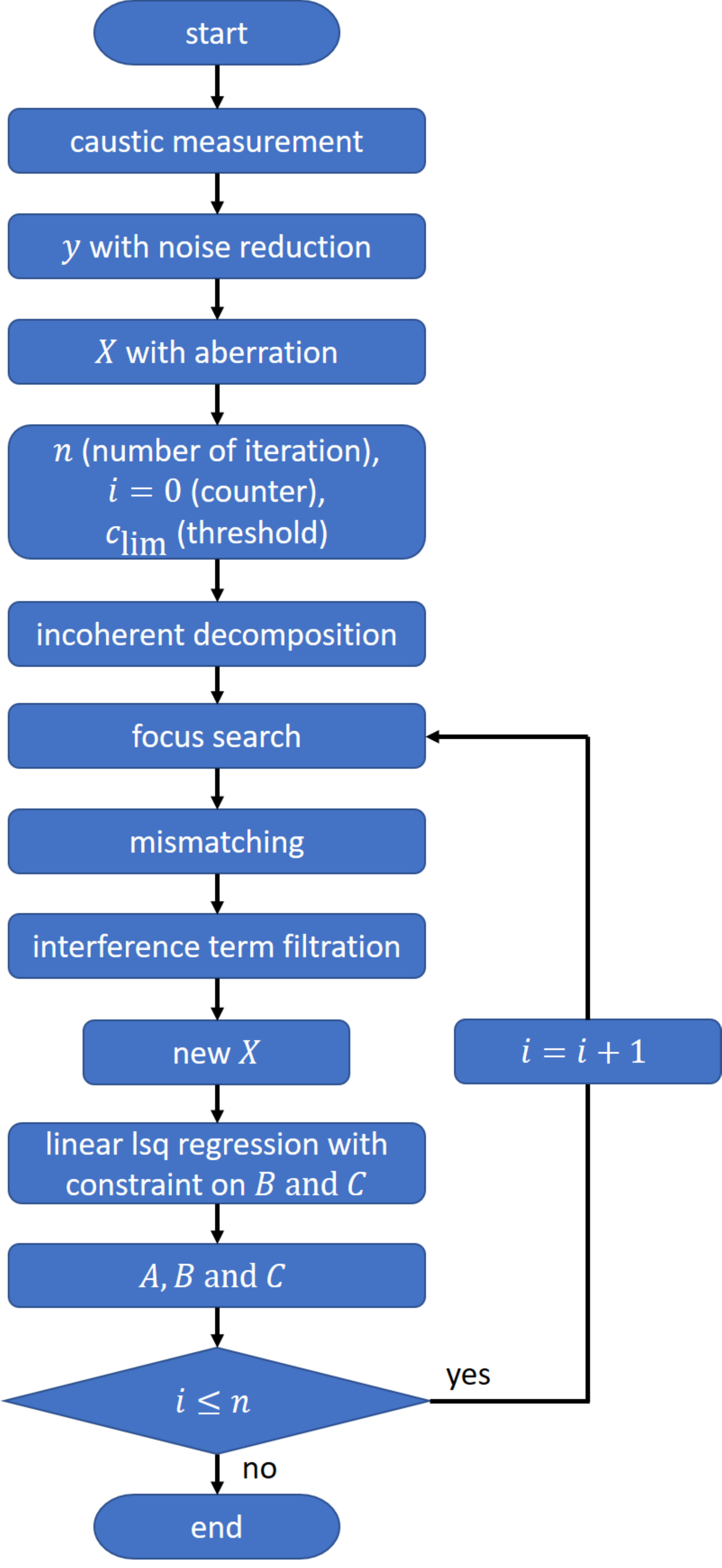}
    \caption{Description of the iterative modal decomposition method. $\A$, $\B$, and $\C$ are the weight factors introduced in Sec.\,\ref{section:3} and \eqref{eq:8}, respectively.}.
    \label{fig:iteration}
\end{figure}
The least square approach presented in Sec.\,\ref{section:3} is only formally linear since the unknown parameters $\A$, $\B$, and $\C$ are connected via \eqref{eq:8}. From $\abs{\gamma_{mn}} \leqslant 1$ it immediately follows that   
\begin{equation}\label{eq:18}
    \mathbf{B}^2+\mathbf{C}^2\leqslant\varrho_m^2\varrho_n^2
\end{equation}
is a necessary condition, where the right-hand side can be easily expressed by the modal weights in $\A$ defined in \eqref{eq:8}. Otherwise, the system tends to reconstruct the detailed structure of intensity profiles (such as interference patterns and aberrations) through over weighted interference terms to achieve a minimal residuum. The overweight of interference terms results in $\abs{\gamma_{mn}}>1$, which makes physically no sense.\\
\begin{figure*}[hbt!] 
    \centering
    \includegraphics[width=\textwidth]{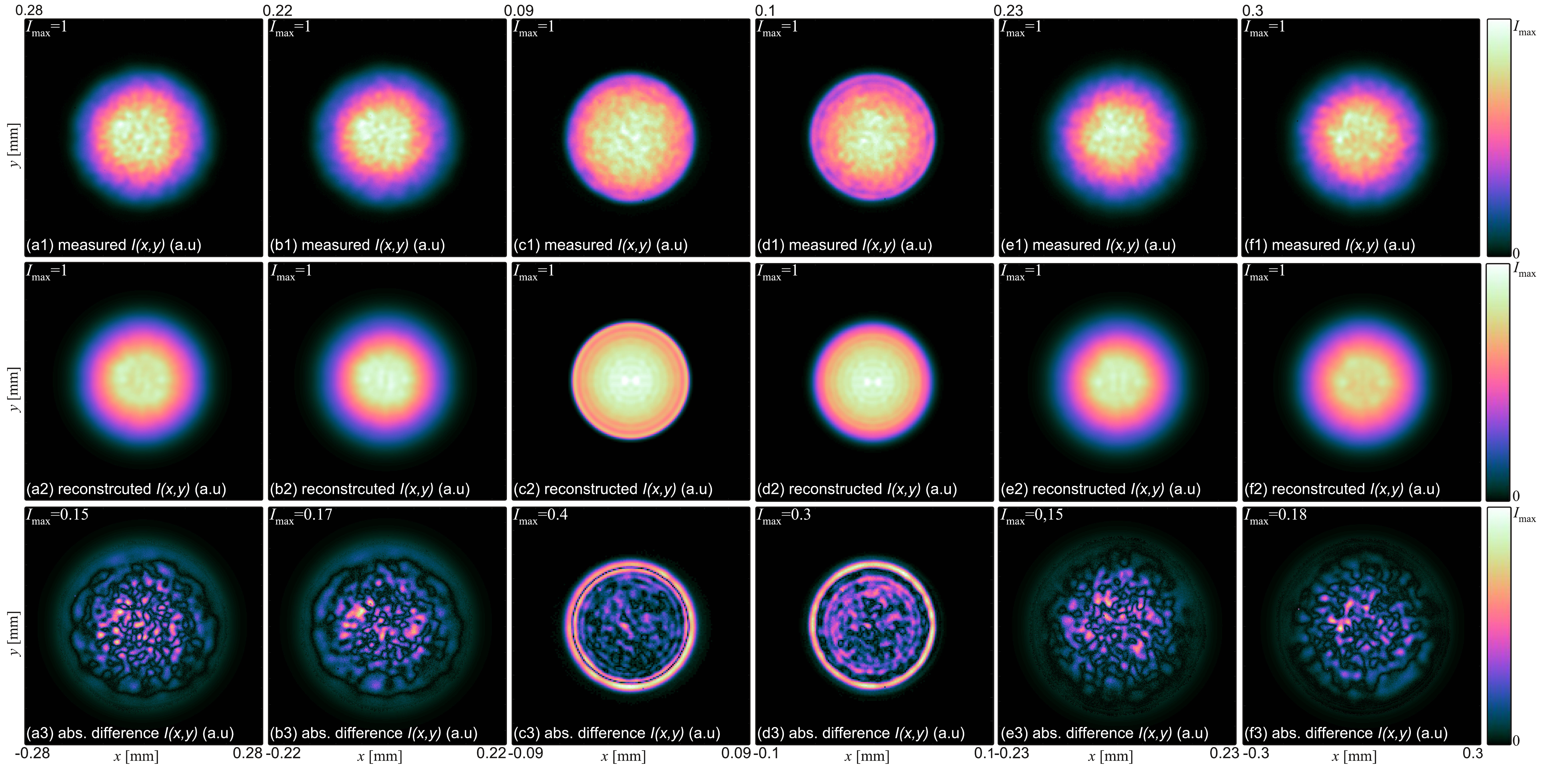}
    \caption{First line: measured intensity distribution of a high power laser beam (\unit[8]{kW}, TruDisk8001). Second line: reconstructed intensity distribution, cf.\,Sec.\,\ref{section:4}\,\ref{sec:selectedExample}. Third line: the absolute value of the difference between measured and reconstructed intensity distributions. The $z$-positions of each column read as: $\unit[1.59]{mm}, \unit[1.22]{mm}, \unit[0.02]{mm}, \unit[-0.01]{mm}, \unit[-1.28]{mm}, \unit[-1.67]{mm}$ ($z=0$ corresponds to fiber end facet). The relative error is defined as the absolute optical power difference between reconstructed and measured profiles divided by the optical power of the corresponding measured profile. For each column from left to right, the relative errors read as: $6\%, 7\%, 19\%, 16\%, 6\%, 6\%$. In this particular case $\clim = 10^{-6}$, cf.\,\eqref{eq:filter}.}
    \label{fig:Intensity1_5_10_11_16_20}
\end{figure*}
In order to consider these parameter interrelations, it would be most accurate to incorporate \eqref{eq:18} in a self-constraint linear regression. However, this is numerically very elaborate and none of the well-established solver algorithms for linear problems could be used. On the other hand, a linear least square regression for fixed constraints is the state of the art and powerful solvers exist, which can be implemented in Python \cite{virtanen2020scipy}. This animated us to design an iterative algorithm for the decomposition. The corresponding workflow is summarized in Fig.\,\ref{fig:iteration}.\\ 
Firstly, a caustic measurement is done in accordance to ISO 11146 \cite{ISO11146}. After the subtraction of the camera noise from the measured data, cf.\,Sec.\,\ref{section:4}\,\ref{sec:section:4A}, the ansatz function matrix $\X$ is created in consideration with aberrations of imaging optics. Since the relevant laser beams considered in this work are rather incoherent (coherence length $<\unit[1]{mm}$, coherent contrast $<0.2$), we start with a completely incoherent decomposition for $\A$. Afterwards, spatial mismatch and focus position are corrected as described previously, cf.\,Sec.\,\ref{section:4}\,\ref{sec:section4C}. The subsequent selection of interference terms according to \eqref{eq:filter} is based on the latest version of $\A$ and the spectrally estimated degree of coherence $\gamma_{mn}$. Next, the constraints for $\B$ and $\C$ have to be fixed according to \eqref{eq:18}, where $\varrho_m, \varrho_n$ are again taken from the latest version of $\A$. However, since $\B$ and $\C$ are separated parameters, we have to state individual constraints and those weaker conditions
\begin{equation}\label{eq:constraint2}
    \mathbf{B}^2\leqslant\varrho_m^2\varrho_n^2, \quad \mathbf{C}^2\leqslant\varrho_m^2\varrho_n^2
\end{equation}
deduced from \eqref{eq:18} are applied. Finally, a constrained linear least square regression is performed which yields the new weights $\A$, $\B$, and $\C$. These can be improved in repeated iterations by taking the current value of $\A$ as input for the mode selection and the definition of the constraints for $\B$ and $\C$. The iteration process terminates after previously defined $n$ times iterations.\\
Generally, the number of considered interference terms fluctuates at the beginning of the iterations. It remains stable or fluctuates periodically after certain iterations are performed. We consider the result as the solution of the system with the given $\clim$. 

\subsection{Discussion of a selected example}\label{sec:selectedExample}
\begin{figure}[h]
    \centering
    \includegraphics[width=\linewidth]{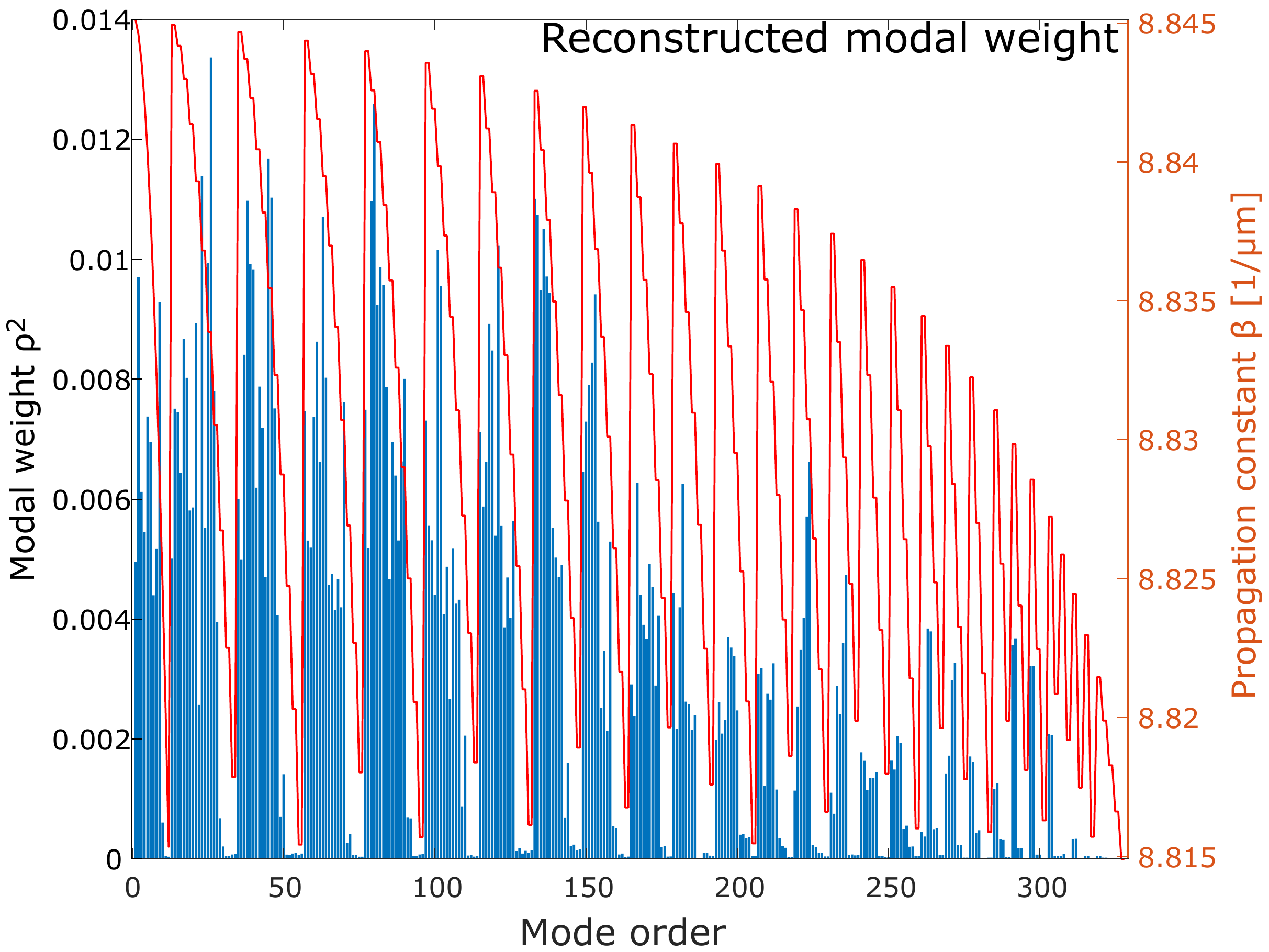}
    \caption{Reconstructed modal weights $\A$ (blue) of the measurement example depicted in Fig.\,\ref{fig:Intensity1_5_10_11_16_20} with corresponding propagation constants $\beta_m$ (red). In this representation the order of the considered 328 LP$_{ij}$ modes reads as: $ij = 01, 02, \dots 0j\textsubscript{max}$, $11, 12, \dots 1j\textsubscript{max}, \dots i\textsubscript{max}j\textsubscript{max}$. Thus, starting from the fundamental mode LP$_{01}$ with maximum $\beta$-value we first increase the radial mode order $j$ up to the respective maximum. Then, the azimuthal order $i$ is increased. Modes with ``even'' and ``odd'' azimuthal dependency \cite{snyder2012optical} follow each other directly.}
    \label{fig:weight1_5_10_11_16_20}
\end{figure}
\begin{figure}[h]
    \centering
    \includegraphics[width=\linewidth]{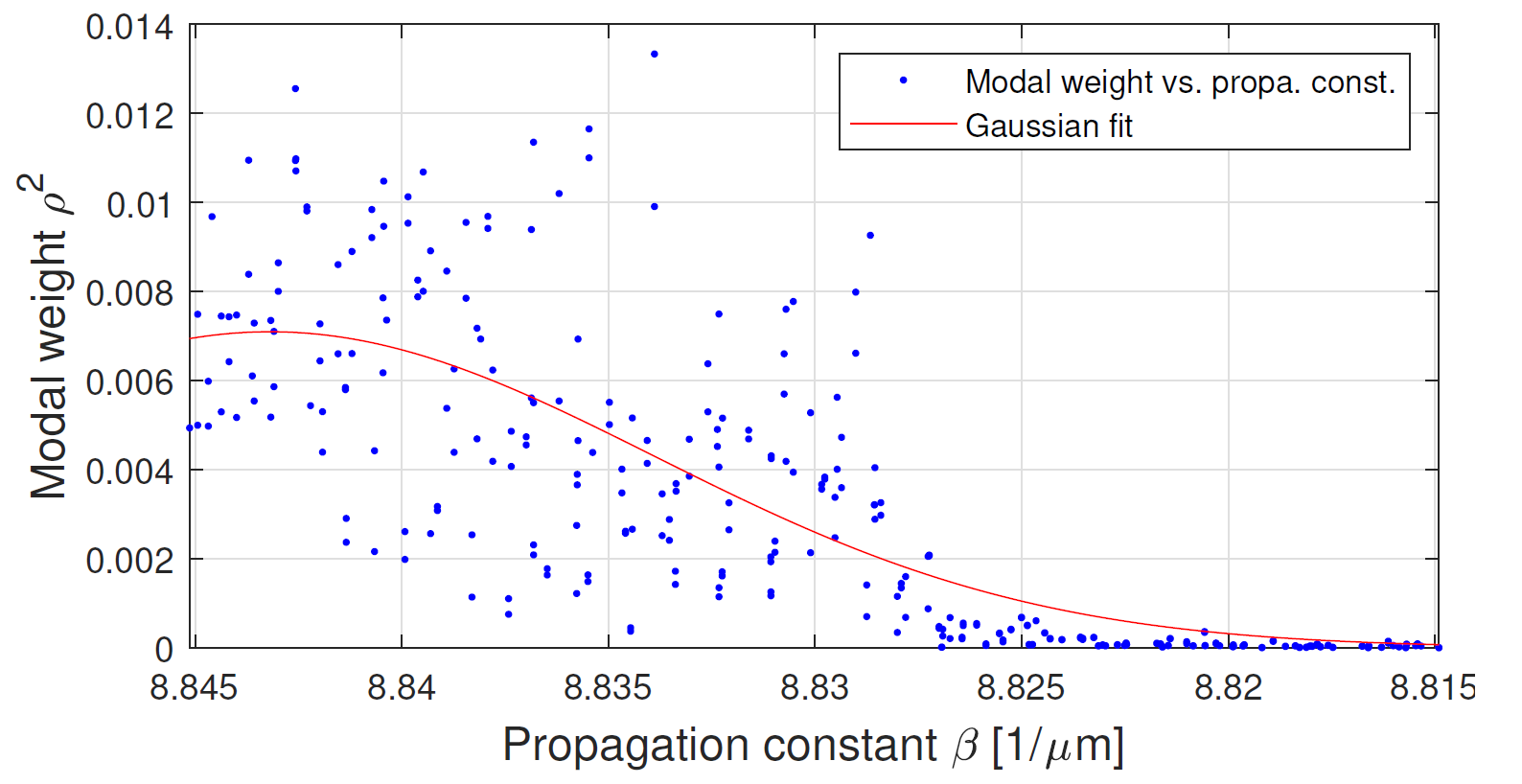}
    \caption{Reconstructed modal weights $\A$ now reordered by decreasing propagation constants $\beta_m$ (blue dots) and the fit result of the reconstructed modal weights of the measurement example depicted in Fig.\,\ref{fig:Intensity1_5_10_11_16_20} (red line).}
    \label{fig:fit curve}
\end{figure}

A reconstruction based on experimental data is shown in Fig.\,\ref{fig:Intensity1_5_10_11_16_20} where 6 intensity measurements along the beam caustic were considered and 4 iterations were performed. The measurement was carried out with a disk laser TruDisk 8001 operating at $\unit[8]{kW}$ and the measurement setup was similar to the setup shown in Fig.\,\ref{fig:aberration} (extra mirrors for attenuation of laser power not shown). The decomposition was performed with a conventional computer in approximately one hour. Due to the neglect of the interference terms, cf.\,Sec.\,\ref{section:4}\,\ref{sec:section4D}, some detailed structures with high-spatial frequencies of the intensity profile can not be reconstructed. According to the third line in Fig.\,\ref{fig:Intensity1_5_10_11_16_20}, we observe a relative stronger deviation in focal area than in far field area. The annular-like deviation, cf.\,(c3) and (d3) of Fig.\,\ref{fig:Intensity1_5_10_11_16_20}, indicates that the size of calculated modes differ slightly from the measurements ($\approx 6\%$ smaller). This might be caused by additional off-axis aberrations or the mismatch of $z$-positions. Generally, the reconstructed  laser beam has a very similar propagation characteristic to the measured laser beam. The beam quality and Rayleigh length of the reconstructed beam differ from the measured values only slightly, see Table\,\ref{tab:measured_reconstructed}.
\begin{table}[]
\centering
\caption{Comparison of the measured and reconstructed beam in terms of the beam quality and the Rayleigh length.}
\begin{tabular}{ccc}
\hline
 & $M^2$ & $z_r \unit[ ]{[\upmu m]}$ \\
\hline
measured & $13.1$ &  $578.2$\\
reconstructed & $12.7$ & $543.5$ \\
\hline
\end{tabular}
  \label{tab:measured_reconstructed}
\end{table}\\
The detailed information of reconstructed modal weights is explicated in Fig.\,\ref{fig:weight1_5_10_11_16_20}. The left vertical axis corresponds to the modal weights (blue bars). On the other hand, the right vertical axis represents the modal propagation constants $\beta_m$ introduced in Eq.\,(\ref{eq:4}) (red line). In this representation the order of the considered 328 LP$_{ij}$ modes reads as: $ij = 01, 02, \dots 0j_{\text{max}}$, $11, 12, \dots 1j_{\text{max}}, \dots i_{\text{max}}j_{\text{max}} $. Thus, starting from the fundamental mode LP$_{01}$ with maximum $\beta$-value, we first increase the radial mode order $j$ up to the respective maximum. Then, the azimuthal order $i$ is increased. Modes with ``even'' and ``odd'' azimuthal dependency \cite{snyder2012optical} follow each other directly. Interestingly, the reconstructed modal power spectrum shows a certain periodic attenuation with increasing radial mode order. In a good approximation, the propagation constants $\beta_m$ act as envelope for the statistically strongly fluctuated modal weights. Modes with low $\beta_m$-values exhibit small relative mode powers $\varrho_m^2$ and, partly, cannot be detected by our metrology. This attenuation behavior is determined by the NA of the in-coupled radiation and the well-known modal bend loss for decreasing $\beta_m$ \cite{schulze2013mode}. We suggest to (roughly) model the dependency $\varrho_m^2\left(\beta_m\right)$ empirically via a simple Gaussian function
\begin{equation}\label{eq:19}
    \rho_{m,\text{fit}}^2 = p_1\exp\left\{-\left[\left(\beta_m-p_2\right)/p_3\right]^2 \right\}.
\end{equation}
In Fig.\,\ref{fig:fit curve}, the reconstructed modal weights corresponding to the decreasing propagation constants in blue dots and the fitting result in a red curve is plotted. Although, the statistical fluctuation is, again, clearly visible, the decreasing behaviour is well described by the fitting curve.\\ 
The position of the centered peak of the Gaussian function $p_2$ is strongly influenced by the near field intensity distribution. In our particular case, the spatial modes fill the light guiding core homogeneously, resulting in a multi-mode flat-top profile, cf.\,(c1) and (d1) in Fig.\,\ref{fig:Intensity1_5_10_11_16_20}. The peak is therefore very close to the maximum propagation constant, thus, $p_2=8.844$. The parameter $p_3$ describes the modal attenuation with decreasing $\beta$-values and depends strongly on the $\NA$ of the laser beam. For this particular case $p_3=0.013$.\\
The approximation of the modal weights distribution into a Gaussian function, cf.\,\eqref{eq:19}, yields sufficiently accurate information about the power spectrum of an investigated fiber and is useful for source modelling if no experimental access to beam's caustic or spectrum is at hand. 

\subsection{Beam shaping example}
The benefit of the presented partial coherent source modelling concept is demonstrated by means of a selected example. Visualization 1 shows the ``flight'' along the focus zone of a designed multi-focus intensity distribution based on a specific processing optics, which is useful for high-power cutting of sheet steel (details not part of this investigation). Local intensity features of this beam shaping example, such as, divergence, edge steepness, uniformity, or interference contrast are available during the propagation. Access to this information is provided by simulating the total intensity as partial coherent sum of weighted transverse fiber modes depending on the propagation distance $I_{\text{tot}}\left(z=z'\right)$, cf.\,\eqref{eq:2}. The modal weight factors $\A,\B$ and $\C$ (cf.\,\eqref{eq:7}) used here origin from the modal decomposition discussed in Fig.\,\ref{fig:Intensity1_5_10_11_16_20} and Fig.\,\ref{fig:weight1_5_10_11_16_20}. Thus, the modal decomposition is the precondition of the shown wave-optical simulation. The employed propagation operator to compute modal fields in different propagation planes $z'$ is based on the angular spectrum of plane waves \cite{goodman2005introduction}.\\

\section{challenges of numerical mode analysis techniques}\label{section:5}
\begin{figure}[h]
    \centering
    \includegraphics[width=\linewidth]{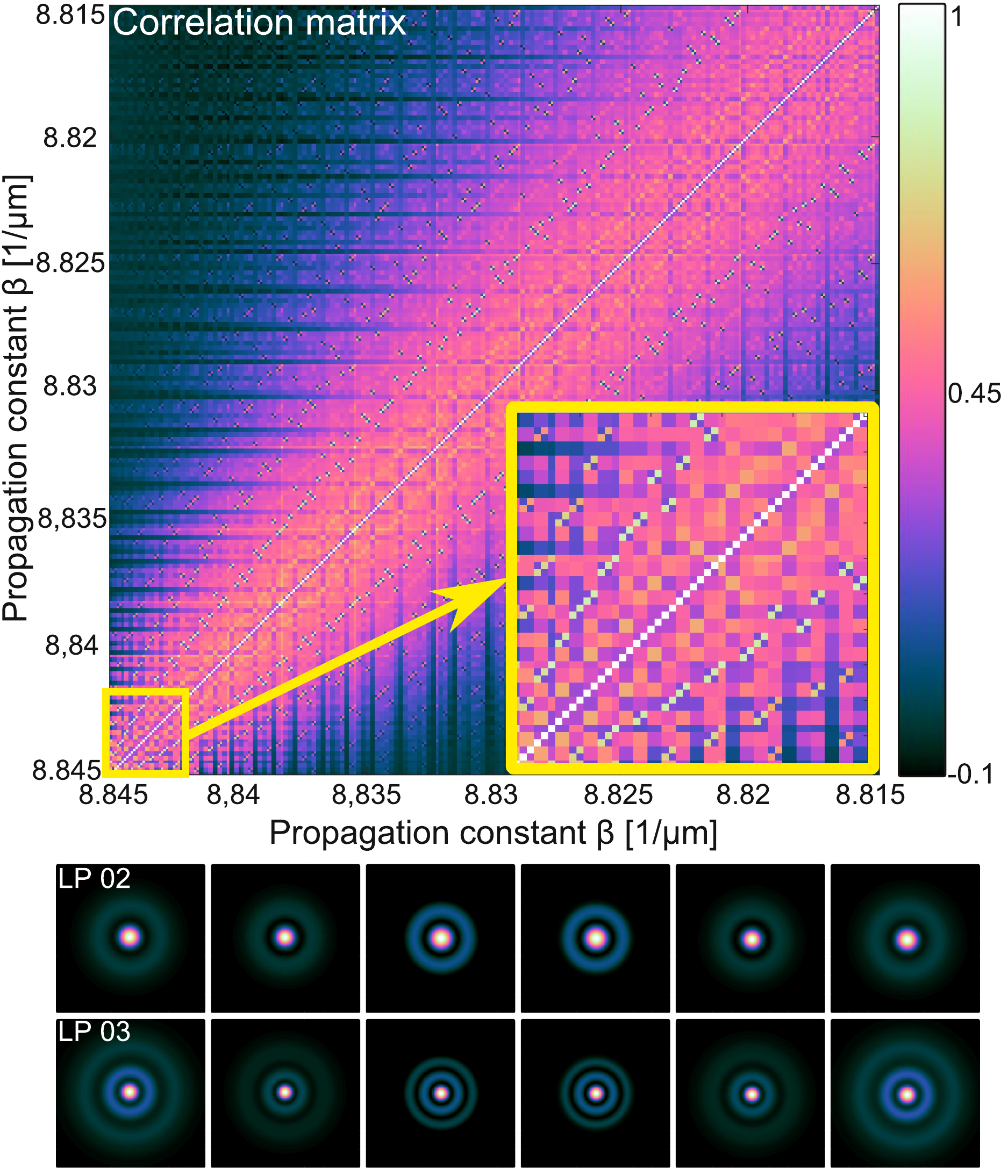}
    \caption{Correlation matrix based on matrix $\mathbf{X}$ (cf.\,\eqref{eq:11}) of modal decomposition introduced in Fig.\,\ref{fig:Intensity1_5_10_11_16_20}. 328 incoherent terms sorted by propagation constants are considered. Bottom: Intensity profiles of LP\textsubscript{02} and LP\textsubscript{03} along the propagation. The corresponding correlation coefficient is 0.85.}
    \label{fig:correlation matrix}
\end{figure}

Considering the modal decomposition results of Sec.\,\ref{section:4}\,\ref{sec:selectedExample} and Fig.\,\ref{fig:Intensity1_5_10_11_16_20}, respectively, we have been able to describe the spatial properties of the light source very precisely. At this point we have to ask ourselves -- and this applies to all related, numerical techniques \cite{shapira2005complete, bruning2013comparative, an2020deep} -- whether these are actually the \textit{physical} mode coefficients of the system, or whether we only have found \textit{one} solution that describes our question in sufficient detail. A quantified statement about the uncertainty of a single mode coefficient cannot be provided here, especially since the number of considered modes is large and the source of errors is multi-dimensional.\\
To understand this ambiguity more thoroughly, the correlation coefficient \cite{schachter1974error} between two arbitrary rows of the matrix $\X$ is considered. In each row of $\X$, not only the intensity profiles, but also the propagation behavior of each mode field is stored, cf.\,\eqref{eq:11}. 
This matrix indicates that although modes pairs are mathematically different (orthogonal), some of their intensity distributions along the propagation might be very similar, resulting in high correlation coefficients ($\geqslant 0.8$), see also example in the bottom of Fig.\,\ref{fig:correlation matrix}. Thus, mode analysis techniques, such as the present one, whose algorithms are based on intensity differences at different propagation distances, cf.\,\eqref{eq:12}, will have difficulties to distinguish certain mode pairs.\\
In analogy to phase diversity concepts known from phase retrieval algorithms \cite{paxman1992joint, mugnier2006phase}, we propose to consider further properties of the multi-mode beam under test. The consideration of diffraction at well-defined obstacles could help numerical mode analysis concepts to distinguish between certain mode groups. For example, diffraction at triangular apertures can be used to unveil orbital angular momentum states of laser radiation \cite{hickmann2010unveiling}. The consideration of such or similar characteristics could help to minimize ambiguities in the present problem. Although the procedure is not demonstrated and only proposed here, we call the concept ``modal diversity.'' \\
Deep learning concepts show a lot of potential to solve mode analysis problems numerically and are recently frequently researched. Some aspects of uncertainty might be mitigated with a proper choice of neural network \cite{an2019numerical,fan2020mitigating}. Although, to our knowledge, these techniques were only applied to few-mode fibers, we believe they are equally beneficial for the analysis of highly multi-mode fiber beams, in particular, if computation time and the minimization of inaccuracies in the determination of mode coefficients is considered. We see particular potential if they could be linked with the above mentioned concept of modal diversity.  
\section{Conclusion}
We presented a modal decomposition approach to analyze the focal field of highly multi-mode optical fibers. It is performed numerically by iterating constrained linear least square regressions with consideration of intermodal degrees of coherence. The concept was applied to multi-kW multi-mode beams with relevance for industrial laser cutting.\\
We have shown that, if experimental access to the beam's caustic and its spectral density distribution is available, an accurate wave-optical description of the light source can be achieved based on the physical modes of the fiber. 
The introduced method can be applied to establish a wave-optical model of highly multi-mode fibers, which is strongly relevant to laser material processing, imaging technologies and fiber characterization. 
\\
Finally, the concept of modal diversity was suggested to mitigate ambiguities occurring especially in the analysis of highly multimodal laser radiation.

\section*{Disclosures}
The authors declare no conflicts of interest.



\bibliography{Lib1}

\begin{thebibliography}{10}
\newcommand{\enquote}[1]{``#1''}

\bibitem{lasermarket1}
{OPTECH CONSULTING}, \enquote{2018 market for lasers and laser systems for
  materials processing,
  \url{http://optech-consulting.com/2018_laser_market_data/},}  (accessed
  online 2019-04-11).

\bibitem{reimann2017three}
W.~Reimann, M.~Dobler, M.~Goede, M.~Schmidt, and K.~Dilger, \enquote{Three-beam
  laser brazing of {zinc-coated} steel,} {\protect\JournalTitle{The
  International Journal of Advanced Manufacturing Technology}} \textbf{90},
  317--328 (2017).

\bibitem{jenne2018high}
M.~Jenne, D.~Flamm, T.~Ouaj, J.~Hellstern, J.~Kleiner, D.~Grossmann,
  M.~Koschig, M.~Kaiser, M.~Kumkar, and S.~Nolte, \enquote{High-quality
  tailored-edge cleaving using aberration-corrected bessel-like beams,}
  {\protect\JournalTitle{Optics Letters}} \textbf{43}, 3164--3167 (2018).

\bibitem{OlsenFl}
F.~O. Olsen, K.~S. Hansen, and J.~S. Nielsen, \enquote{Multibeam fiber laser
  cutting,} {\protect\JournalTitle{Journal of Laser Applications}} \textbf{21},
  133--138 (2009).

\bibitem{goppold2014experimental}
C.~Goppold, K.~Zenger, P.~Herwig, A.~Wetzig, A.~Mahrle, and E.~Beyer,
  \enquote{Experimental analysis for improvements of process efficiency and cut
  edge quality of fusion cutting with 1 $\mu$m laser radiation,}
  {\protect\JournalTitle{Physics Procedia}} \textbf{56}, 892--900 (2014).

\bibitem{flamm2019beam}
D.~Flamm, D.~G. Grossmann, M.~Jenne, F.~Zimmermann, J.~Kleiner, M.~Kaiser,
  J.~Hellstern, C.~Tillkorn, and M.~Kumkar, \enquote{Beam shaping for ultrafast
  materials processing,} in \emph{Laser Resonators, Microresonators, and Beam
  Control XXI,} , vol. 10904 (International Society for Optics and Photonics,
  2019), p. 109041G.

\bibitem{hugel2009laser}
H.~H{\"u}gel and T.~Graf, \emph{Laser in der Fertigung}, vol.~2 (Springer,
  2009).

\bibitem{Wolf:82}
E.~Wolf, \enquote{New theory of partial coherence in the space--frequency
  domain. {Part I: Spectra} and cross spectra of steady-state sources,}
  {\protect\JournalTitle{J. Opt. Soc. Am.}} \textbf{72}, 343--351 (1982).

\bibitem{Wolf:86}
E.~Wolf, \enquote{New theory of partial coherence in the space-frequency
  domain. {Part II: Steady-state} fields and higher-order correlations,}
  {\protect\JournalTitle{J. Opt. Soc. Am. A}} \textbf{3}, 76--85 (1986).

\bibitem{snyder2012optical}
A.~W. Snyder and J.~Love, \emph{Optical waveguide theory} (Springer Science \&
  Business Media, 2012).

\bibitem{flamm2013modal}
D.~Flamm, \enquote{The modal transmission function of optical fibers,} Ph.D.
  thesis, Friedrich Schiller University Jena (2013).

\bibitem{shapira2005complete}
O.~Shapira, A.~F. Abouraddy, J.~D. Joannopoulos, and Y.~Fink, \enquote{Complete
  modal decomposition for optical waveguides,} {\protect\JournalTitle{Physical
  review letters}} \textbf{94}, 143902 (2005).

\bibitem{kaiser2009complete}
T.~Kaiser, D.~Flamm, S.~Schr{\"o}ter, and M.~Duparr{\'e}, \enquote{Complete
  modal decomposition for optical fibers using {CGH-based} correlation
  filters,} {\protect\JournalTitle{Optics Express}} \textbf{17}, 9347--9356
  (2009).

\bibitem{bruning2013comparative}
R.~Br{\"u}ning, P.~Gelszinnis, C.~Schulze, D.~Flamm, and M.~Duparr{\'e},
  \enquote{Comparative analysis of numerical methods for the mode analysis of
  laser beams,} {\protect\JournalTitle{Applied Optics}} \textbf{52}, 7769--7777
  (2013).

\bibitem{forbes2016creation}
A.~Forbes, A.~Dudley, and M.~McLaren, \enquote{Creation and detection of
  optical modes with spatial light modulators,} {\protect\JournalTitle{Advances
  in Optics and Photonics}} \textbf{8}, 200--227 (2016).

\bibitem{nicholson2008spatially}
J.~Nicholson, A.~D. Yablon, S.~Ramachandran, and S.~Ghalmi, \enquote{Spatially
  and spectrally resolved imaging of modal content in large-mode-area fibers,}
  {\protect\JournalTitle{Optics Express}} \textbf{16}, 7233--7243 (2008).

\bibitem{tillkorn2018anamorphic}
C.~Tillkorn, A.~Heimes, D.~Flamm, S.~Dorer, T.~Beck, J.~Hellstern,
  F.~Marschall, and C.~Lingel, \enquote{Anamorphic beam shaping for efficient
  laser homogenization: Methods and high power applications,} in \emph{Laser
  Resonators, Microresonators, and Beam Control XX,} , vol. 10518
  (International Society for Optics and Photonics, 2018), p. 105181I.

\bibitem{flusberg2005fiber}
B.~A. Flusberg, E.~D. Cocker, W.~Piyawattanametha, J.~C. Jung, E.~L. Cheung,
  and M.~J. Schnitzer, \enquote{Fiber-optic fluorescence imaging,}
  {\protect\JournalTitle{Nature methods}} \textbf{2}, 941 (2005).

\bibitem{borhani2018learning}
N.~Borhani, E.~Kakkava, C.~Moser, and D.~Psaltis, \enquote{Learning to see
  through multimode fibers,} {\protect\JournalTitle{Optica}} \textbf{5},
  960--966 (2018).

\bibitem{ploschner2015seeing}
M.~Pl{\"o}schner, T.~Tyc, and T.~{\v{C}}i{\v{z}}m{\'a}r, \enquote{Seeing
  through chaos in multimode fibres,} {\protect\JournalTitle{Nature Photonics}}
  \textbf{9}, 529 (2015).

\bibitem{jacobs2002optical}
I.~Jacobs, \enquote{Optical fiber communication technology and system
  overview,} {\protect\JournalTitle{FIBER OPTICS HANDBOOK}}  (2002).

\bibitem{otto2013improved}
H.-J. Otto, F.~Jansen, F.~Stutzki, C.~Jauregui, J.~Limpert, and A.~Tunnermann,
  \enquote{Improved modal reconstruction for spatially and spectrally resolved
  imaging,} {\protect\JournalTitle{Journal of lightwave technology}}
  \textbf{31}, 1295--1299 (2013).

\bibitem{cohen1998generalization}
L.~Cohen, \enquote{The generalization of the {Wiener-Khinchin} theorem,} in
  \emph{Proceedings of the 1998 IEEE International Conference on Acoustics,
  Speech and Signal Processing, ICASSP'98 (Cat. No. 98CH36181),} , vol.~3
  (IEEE, 1998), pp. 1577--1580.

\bibitem{born2013principles}
M.~Born and E.~Wolf, \emph{Principles of optics: electromagnetic theory of
  propagation, interference and diffraction of light} (Elsevier, 2013).

\bibitem{hodgson2005laser}
N.~Hodgson and H.~Weber, \emph{Laser Resonators and Beam Propagation:
  Fundamentals, Advanced Concepts, Applications}, vol. 108 (Springer, 2005).

\bibitem{ISO11146}
{International Organization for Standardization}, \enquote{Lasers and
  laser-related equipment — {Test} methods for laser beam widths, divergence
  angles and beam propagation ratios — {Part 1: Stigmatic and simple
  astigmatic beams},} Standard, International Organization for Standardization,
  Geneva, CH (2005).

\bibitem{merx2020TBS}
S.~Merx, J.~Stock, F.~Widiasari, and H.~Gross, \enquote{Beam characterization
  by phase retrieval solving the transport-of-intensity-equation,}
  {\protect\JournalTitle{Optics Express (accepted),
  http://dx.doi.org/10.1364/OE.394633}}  (2020).

\bibitem{virtanen2020scipy}
P.~Virtanen, R.~Gommers, T.~E. Oliphant, M.~Haberland, T.~Reddy, D.~Cournapeau,
  E.~Burovski, P.~Peterson, W.~Weckesser, J.~Bright \emph{et~al.},
  \enquote{Scipy 1.0: fundamental algorithms for scientific computing in
  python,} {\protect\JournalTitle{Nature methods}} \textbf{17}, 261--272
  (2020).

\bibitem{schulze2013mode}
C.~Schulze, A.~Lorenz, D.~Flamm, A.~Hartung, S.~Schr{\"o}ter, H.~Bartelt, and
  M.~Duparr{\'e}, \enquote{Mode resolved bend loss in few-mode optical fibers,}
  {\protect\JournalTitle{Optics Express}} \textbf{21}, 3170--3181 (2013).

\bibitem{goodman2005introduction}
J.~W. Goodman, \emph{Introduction to Fourier optics} (Roberts and Company
  Publishers, 2005).

\bibitem{an2020deep}
Y.~An, L.~Huang, J.~Li, J.~Leng, L.~Yang, and P.~Zhou, \enquote{Deep
  learning-based real-time mode decomposition for multimode fibers,}
  {\protect\JournalTitle{IEEE Journal of Selected Topics in Quantum
  Electronics}}  (2020).

\bibitem{schachter1974error}
J.~Schachter, \enquote{An error in error analysis 1,}
  {\protect\JournalTitle{Language learning}} \textbf{24}, 205--214 (1974).

\bibitem{paxman1992joint}
R.~G. Paxman, T.~J. Schulz, and J.~R. Fienup, \enquote{Joint estimation of
  object and aberrations by using phase diversity,} {\protect\JournalTitle{JOSA
  A}} \textbf{9}, 1072--1085 (1992).

\bibitem{mugnier2006phase}
L.~M. Mugnier, A.~Blanc, and J.~Idier, \enquote{Phase diversity: a technique
  for wave-front sensing and for diffraction-limited imaging,}
  {\protect\JournalTitle{Advances in Imaging and Electron Physics}}
  \textbf{141}, 1--76 (2006).

\bibitem{hickmann2010unveiling}
J.~Hickmann, E.~Fonseca, W.~Soares, and S.~Ch{\'a}vez-Cerda, \enquote{Unveiling
  a truncated optical lattice associated with a triangular aperture using
  light’s orbital angular momentum,} {\protect\JournalTitle{{Physical review
  letters}}} \textbf{105}, 053904 (2010).

\bibitem{an2019numerical}
Y.~An, J.~Li, L.~Huang, L.~Li, J.~Leng, L.~Yang, and P.~Zhou,
  \enquote{Numerical mode decomposition for multimode fiber: From
  multi-variable optimization to deep learning,} {\protect\JournalTitle{Optical
  Fiber Technology}} \textbf{52}, 101960 (2019).

\bibitem{fan2020mitigating}
X.~Fan, F.~Ren, Y.~Xie, Y.~Zhang, J.~Niu, J.~Zhang, and J.~Wang,
  \enquote{Mitigating ambiguity by deep-learning-based modal decomposition
  method,} {\protect\JournalTitle{Optics Communications}} p. 125845 (2020).

\end{thebibliography}

\bigskip
\noindent


\bibliographyfullrefs{Lib1}


\ifthenelse{\equal{\journalref}{aop}}{%
\section*{Author Biographies}
\begingroup
\setlength\intextsep{0pt}
\begin{minipage}[t][6.3cm][t]{1.0\textwidth} 
  \begin{wrapfigure}{L}{0.25\textwidth}
    \includegraphics[width=0.25\textwidth]{john_smith.eps}
  \end{wrapfigure}
  \noindent
  {\bfseries John Smith} received his BSc (Mathematics) in 2000 from The University of Maryland. His research interests include lasers and optics.
\end{minipage}
\begin{minipage}{1.0\textwidth}
  \begin{wrapfigure}{L}{0.25\textwidth}
    \includegraphics[width=0.25\textwidth]{alice_smith.eps}
  \end{wrapfigure}
  \noindent
  {\bfseries Alice Smith} also received her BSc (Mathematics) in 2000 from The University of Maryland. Her research interests also include lasers and optics.
\end{minipage}
\endgroup
}{}

\end{document}